\documentstyle[12pt]{article}   
\begin{document}

\begin{center}
{\large{\bf
A semiclassical theory of quantum noise in open chaotic systems}}
\end{center}

\begin{center}
{\bf
Bidhan Chandra Bag, Shanta Chaudhuri, \\
Jyotipratim Ray Chaudhuri and Deb Shankar Ray \\
Indian Association for the Cultivation of Science \\
Jadavpur, Calcutta 700 032, INDIA }
\end{center}

\begin{abstract}
We consider the quantum evolution of  classically chaotic systems in contact  
with  surroundings. Based on $\hbar$-scaling of an equation for 
time evolution of the Wigner's quasi-probability 
distribution function in presence of 
dissipation and thermal diffusion we derive a semiclassical equation
for quantum fluctuations. This
identifies an early regime of evolution dominated by fluctuations in the 
curvature of the potential due to classical chaos and dissipation. 
A stochastic treatment 
of this classical fluctuations leads us to a Fokker-Planck 
equation which is reminiscent
of Kramers' equation for thermally activated processes. 
This reveals an interplay of three aspects of evolution of quantum noise
in weakly dissipative open systems; the reversible 
Liouville flow, the 
irreversible
chaotic diffusion which is characteristic of the system itself, and 
irreversible dissipation induced by the external reservoir. It has been 
demonstrated that in the dissipation-free case a competition between Liouville
flow in the contracting direction of phase space and chaotic diffusion sets
a critical width in the Wigner function for quantum fluctuations. We also 
show how the initial quantum noise gets amplified by classical chaos and
ultimately equilibrated under the influence of dissipation. We establish
that there exists a critical limit to the expansion of phase space. The limit
is determined by chaotic diffusion and dissipation. Making use of appropriate
quantum-classical correspondence we verify the semiclassical analysis by the
fully quantum simulation in a chaotic quartic oscillator.
\end{abstract}

\newpage

\begin{center}
{\bf{I. \hspace{0.2cm}Introduction}}
\end{center}

\vspace{0.5cm}
 
Dissipation in quantum dynamical system has been one of the most                 
intriguing issues in physics. Although much of our understanding 
of dissipative linear systems [1] forms a well-developed body of 
literature by now, the interplay
of nonlinearity and dissipation [4-11] has drawn major attention in recent years, 
particularly in the problems 
relating to macroscopic quantum tunneling [2], multiphoton 
dissociation dynamics of molecules [3], 
quantum decoherence [4] etc. The class of 
nonlinear systems which are classically chaotic offers a good opportunity in 
this context, to understand the role of dissipation in quantum-classical
correspondence [5]. The coupling of the system 
with its surroundings induces 
exchange of energy between them resulting in 
dissipation of energy of the system.
This openness also imparts classicality in the quantum system to the extent
that quantum localization in a classically chaotic system gets suppressed.
The subject was analyzed early by Dittrich and Graham [5] on the 
basis of a quantized standard map. In an another issuse Cohen [5] considerd
the problem of localization in the quantum kicked rotator model leading to
nontrivial dynamical correlations where quantum chaos  has an
interesting bearing on the destruction of coherence. Quantum decoherence
in the context of  quantum-classical correspondence in several model 
systems, e. g. , circle and stadium billiards has been the subject of 
further investigation [4].

More recently a number of numerical experiments have demonstrated 
[6-8] that the 
initial growth of quantum variances of the canonical dynamical variables
such as, position or momentum for a classically chaotic 
trajectory is exponential in 
nature. This has been identified as a typical signature of classical chaos on 
a generic quantum dynamical feature, or more precisely, a semiclassical 
manifestation of classical chaos. 
The manifestation of the classical chaotic spreading in the initial phase of 
a dissipative quantum dynamics has been treated recently in a general and 
elegant manner by Pattanayak and Brumer [11].
We have shown earlier that the fluctuation in the curvature
of the potential [8-10] 
in the dynamical system, is amenable to a stochastic 
description in terms of the theory of multiplicative noise. The 
origin of classical instability and early divergence of quantum variances can be 
traced back to the correlation functions of fluctuations of the 
curvature of the classical
potential. The present study focuses on two specific issues ; firstly
we address the problem of  
evolution of quantum noise in an open system at the 
semiclassical level which identifies the interplay of three distinct aspects of evolution, 
e. g. , (a) deterministic Liouville flow (b) irreversible chaotic diffusion and 
(c) irreversible dissipation due to the external surroundings. Second, 
we explore the role of dissipation in the ultimate equilibration of the 
quantum noise in presence of classical chaos.
Based on Wigner's quantum-classical correspondence
we have derived the appropriate Fokker-Planck equation where the drift and 
diffusion terms have their origin in the dynamical properties of the 
fluctuations of the curvature of the classical potential and dissipation 
due to the coupling of the chaotic system 
to the surroundings. Our results show how the initial quantum noise gets 
amplified by chaotic diffusion and ultimately settles down to equilibrium with
the passage of time under the influence of dissipation. In the 
dissipation-free case, an interplay of reversible Liouville flow and chaotic diffusion
sets a limit on the width of Wigner function undergoing evolution. We also
establish that there exists a critical limit to the expansion of phase
space determined by dissipation and chaotic diffusion.
A detailed analysis of classical and 
quantum mechanical calculations on a driven quartic double-well model has been 
carried out for numerical verification of our semiclassical analysis.
We point out that the dissipative quantum dynamics of a similar 
system has recently been treated in a technically related approach by Dittrich,
Oelschlaegel and H\"anggi [12].

The organization of the paper is as follows ; In Sec. II we provide  
a background for quantum evolution of a classically chaotic system in presence 
of dissipation and thermal diffusion in the density matrix picture. A 
c-number  formulation of the equation is described in Sec. III in 
terms of Wigner's quasi-classical probability function. $\hbar$-scaling
of the equation of motion is then carried out to derive a semiclassical
equation which identifies an early stage of evolution dominated by 
dissipation and curvature of the potential. Based on a cumulant expansion in 
$\alpha \tau_c$ where $\alpha$ is the strength and $\tau_c$ is the correlation 
time of fluctuations a Fokker-Planck equation is formulated. In Sec. IV we 
identify three distinct aspects of evolution in terms of a generic model 
driven 
quartic oscillator and analyze their interplay in two distinct situations.
In Sec. V, the theoretical results have been compared with detailed fully
quantum mechanical calculations. The paper is concluded in Sec. VI with a 
summary of the main results.

\vspace{0.5cm}

\begin{center}
{\bf
II. \hspace{0.2cm}Quantum dynamics in presence of dissipation and thermal 
diffusion}
\end{center}

\vspace{0.5cm}

To study [1] 
the evolution of a quantum system in presence of weak dissipation 
and thermal diffusion from the reservoir modes we first consider 
the Hamiltonian of an N-degree-of-freedom system $H_0$.
\begin{equation}
H_0 = \sum_{i=1}^N \frac{p_i^2}{2 m_i} +V(\{x_i\}) \; \;,
\; \; i = 1 \cdots N
\end{equation}

\noindent
where $\{x_i, p_i\}$ represents the coordinates and 
momenta of the N-degree-of-freedom system.

The bare system is now coupled to an environment modeled by a reservoir of 
harmonic oscillator modes. The generation of quantum dynamics is given by 
the overall Hamiltonian operator for the system and the environment and the 
coupling
\begin{equation}
H=H_0+\hbar \sum_i \omega_i b_i^\dagger b_i + \hbar \sum_i \left[ K(\omega_i)b_i
+K^*(\omega_i) b_i^\dagger \right] x \; \; \; ,
\end{equation}
where $x$ and $p$ are position and momentum operator corresponding to a selected 
degree of freedom of the system;       
$b_i (b_i^\dagger)$ denotes the annihilation (creation) operator 
of the harmonic oscillator bath modes. The second and third terms correspond 
to reservoir modes and their linear coupling to the chaotic system. 
$K(\omega_i)$ is a c-function.

It is convenient to invoke the rotating wave approximation (RWA) so that one can 
use a symmetric coupling of the type 
$(b_i a^\dagger + b_i^\dagger a)$, where $a$ and $a^\dagger$, are 
annihilation and creation operators corresponding to the system operator 
co-ordinate $x=\frac{1}{\sqrt{2 m \omega}}(a+a^\dagger)$;
$\omega$ refers to the frequency of the harmonic 
oscillator on the basis of which quantum calculations are performed as 
described in the latter part of the text. 

Appropriate elimination of reservoir modes in the usual way, using Born and 
Markov approximations leads us to the following reduced density matrix 
equation for the evolution of the system [1]
\begin{equation}
\frac{d \rho}{dt} = - \frac{i}{\hbar} [H_0, \rho] + \frac{\gamma}{2}
(2 a \rho a^\dagger - a^\dagger a \rho- \rho a^\dagger a)
+D(a^\dagger \rho a+ a\rho a^\dagger -a^\dagger a \rho - \rho a a^\dagger)
\end{equation}

Here the spectral density function of the reservoir is replaced by a continuous 
density $g(\omega)$ and we denote Boltzmann constant by $k$ and $\gamma > 0$      
is the limit of $2 \pi |K(\omega)|^2 g(\omega)/\omega$
as $\omega \rightarrow 0_+$ and is assumed to be finite.     
$\gamma$ is the relaxation or dissipation rate, $D(=\bar{n} \gamma)$
is the diffusion coefficient and
$\bar{n}(= [exp \left( \frac{\hbar \omega}{k T} \right)-1]^{-1})$
is the average thermal photon number of the reservoir. The terms
analogous to Lamb and Stark shifts have been neglected.

The first term in Eq. (3) corresponds to the
dynamical motion of the system that
generates Liouville flow. The terms containing $\gamma$ 
arise due to the interaction with
the surroundings. 
The first term implies the loss of energy from the system to the
reservoir, while the last term indicates the diffusion of fluctuations of the 
reservoir modes into the 
system of interest. The last term is responsible for 
quantum decoherence processes. In the limit 
$T \rightarrow 0$ the diffusion term in Eq. (3)
vanishes, whereby the system decays primarily due to purely quantum noise.

We now make a few remarks on the approximations involved in Eq. (3)
and its range of applicability in numerical simulation of full quantum
dynamics as carried out in Sec. V.

(i) Since the system-reservoir dynamics as governed by the operator
master Eq. (3) is based on Born-Markov approximation [the correlation time of 
the reservoir must be very short (Markov) for the interaction between
the system and the reservoir to be sufficiently small (Born/weak coupling)],
the underlying stochastic process due to the reservoir is {\it Markovian by 
construction}. We mention here that this has nothing to do with the stochasticity
due to classical chaos which results in fluctuations of the curvature of the
of the potential. This fluctuation has to be taken care of at a {\it non-Markovian}
level of description because of its finite (but short) correlation time.
While we note that there is a vast body of literature in condensed matter and 
chemical physics dealing with finite response time of the reservoir, which
results in frequency dependence of friction coefficiant $\gamma$, these and 
the related aspects of dissipative dynamics are outside the scope of Eq. (3).
Our approach here is similar to that of Graham et. al. [5] in this regard.

(ii) Eq. (3) because of Born approximation
is valid for weak damping case. It is necessary to take care of 
this limitation by choosing small values of $\gamma$ while varying it in
carrying out numerical simulation
of the quantum master equation Eq. (3). 

(iii) It must also be noted that $\gamma$ and $D$ terms in Eq. (3) 
are valid if the system operators pertain to a harmonic
oscillator. When the system is nonlinear, as the present case, the usual
practice is to add the additional contribution $-i[H_{non}, \rho]$ to the 
master equation [in the language of Fokker-Planck description this commutator, in 
general, contributes higher (third or more) order derivatives of the distribution]
and to assume that the dissipative terms remain unaffected by the 
addition of commutator term, $H_{non}$ being the nonlinear part of the 
Hamiltonian. The validity of this assumption was examined [3] earlier by Haake
et. al. and also by us. It is now known that this assumption is quite satisfactory 
within the perview of weak damping and/or high temperature limit.

We note that Eq. (3) is a popular form of the operator master equation, as 
derived by Louisell [1], which is widely used in quantum optics. This 
equation had also been applied earlier by Graham et. al. [5] in the treatment
of dissipative standard map and related problems of chaotic dynamics by others
[13] . The correlation between different forms of operator master equations has been
reviewed in Ref [2]. All of them, however, are not well-suited for numerical
simulations. Eq. (3) suits this purpose well. We shall return to this issue
in Sec. V to verify the theoretical propositions. 

\newpage

\begin{center}
\bf{III. \hspace{0.2cm}Semiclassical dynamics} 
\end{center}

\vspace{0.5cm}

\begin{center}
{\bf A. $\hbar$-scaling and the semiclassical equation}
\end{center}

Our next task is 
to go over from a full quantum operator problem to an equivalent `classical' or more 
appropriately c-number problem  described by the same Hamiltonian (2).
Over the years the standard strategy of analysis of 
quantum-classical correspondence is the quasi-classical distribution function 
of Wigner[14], which is defined in phase space $\{x_i, p_i\}$ as follows;
\begin{eqnarray*}
W(\{x_i\}, \{p_i\}) & = & \frac{1}{(\hbar \pi)^n} \int \cdots \int
d \xi_1 \cdots d \xi_n \psi^\dagger (\{x_i+\xi_i\}) \\
& & 
\times \hspace{0.05cm} \psi(\{x_i-\xi_i\}) \exp\left[ \frac{2i}{\hbar} 
\left( \sum_{i} p_i \xi_i \right) \right] \; \; \; ,
\end{eqnarray*}
where $\psi(x)$ refers the quantum wave function of the 
N-degree-of-freedom system.

The time evolution of Wigner function W of the dynamical system in presence 
of dissipation is now given by, 
\begin{eqnarray}
\frac{\partial W}{\partial t} & = & \sum_{i=1}^N
\left[ -\frac{p_i}{2 m_i} \frac{\partial W}{\partial x_i}
+ \left( \frac{\partial V}{\partial x_i} \right) 
\frac{\partial W}{\partial p_i}
\right] \nonumber \\
& + & \sum_
{\begin{array}{c}
n_1+n_3+\cdots +n_N \; \; is \; \; odd \; \; and \; >1  \\
\end{array}}
\left(\frac{\partial^{n_1+ \cdots n_N}V}{\partial x_1^{n_1} \cdots
\partial x_N^{n_N}} \right)  \frac{\left( \frac{\hbar}{2 i}\right)^
{n_1+ \cdots +n_N-1}}{n_1! \cdots n_N!}  \nonumber \\
& & \times \hspace{0.05cm} \frac{\partial^{n_1+ \cdots +N_N}}{\partial p_1^{n_1} \cdots \partial
p_N^{n_N}}W + 2 \gamma \frac{\partial}{\partial p} p W + D \frac
{\partial^2 W}{\partial p^2} \; \; \; .
\end{eqnarray}

The first term is the usual Poisson bracket which generates the Liouville
flow. Both the Poisson bracket and the 
higher derivative terms result from an
expansion of the Moyal bracket on the basis 
of an analytic $V(x)$. The last two 
terms are due to dissipation and diffusion induced by the external reservoir.
It is important to note 
that the failure of correspondence between classical and quantum dynamics is 
predominantly due to higher derivative terms [14] 
which make their presence felt
roughly beyond the Ehrenfest regime.

The above equation (4) is a full quantum mechanical equation as derived by
Caldeira and Leggett [2]. The primary reasons for choosing Eq.(4) as our 
starting point for semiclassical analysis are: (i) in deriving Eq.(4) the rotating 
wave approximation (RWA) in the system-reservoir coupling has not been made.
Had RWA been used Eq.(4) would have contained additional contribution of 
terms such as
$\frac{\partial xW}{\partial x}$ and $\frac{\partial^2 W}{\partial x^2}$.
(ii) Eq.(4) is also free from Born approximation (or weak coupling 
approximation) ensuring that the theory is valid even in strong damping limit
in contrast to Eq.(3) whose validity is restricted only to the weak damping
regime. For a comparison over the entire range of dissipation one needs, 
however, other kinds of master equation which are free from weak coupling. 
Unfortunately, as we have already pointed out, most of them are not 
well-suited for numerical implementation.
(iii) Eq.(4) reaches the correct classical limit when $\hbar \rightarrow 0$ 
and
$D$ reduces to thermal diffusion coefficient in the high temperature limit. 
Eq.(4) is thus likely to serve as a good description in our semiclassical analysis.

In the next two steps we invoke the symplectic structure of the Hamiltonian
dynamics by defining

\begin{equation}
z_i =
\left\{ \begin{array}{ll}
x_i & \; \; {\rm for} \; i=1 \cdots N \\
p_{i-N} & \; \; {\rm for} \; i=N+1, \cdots 2N  \; \; ,
\end{array} \right.
\end{equation}

\noindent
and introduce the scaling of $z_i$ in analogy to van Kampen's 
$\Omega$ - expansion as

\begin{equation}
z_i=z_i(t)+\hbar^{1/2} {\bf \eta}_i
\end{equation}

where
\begin{eqnarray*}
\eta_i & = & \mu_i \,
\; \; \; \; \; \; {\rm for}\; \; i=1, \cdots N \nonumber \\
& = & \nu_{i-N} \; \; \; {\rm for}\; \; i=N+1 \cdots 2N .
\end{eqnarray*}

\noindent
$\eta$-s refer to quantum noise variables in co-ordinate ($\mu_i$) and 
momentum ($\nu_i$). The equation of motion for quantum fluctuation distribution
function $\phi$ ($\eta, t$) is given by (for details we refer to [8]) 

\begin{equation}
\frac{\partial \phi}{\partial t} = [ -{\bf F}(t) \cdot {\bf \nabla} 
+ 2 N \gamma] \phi \; \; \; ,
\end{equation}

where ${\bf F}(t) = \underline{J}(t) {\bf \eta} - 2 \gamma \underline{K} {\bf \eta}$ ;
${\bf \nabla}$ refers to 
differentiation with respect to the components of ${\bf \eta}$. \underline{K}
is a $2N \otimes 2N$ matrix defined as $k_{ij} = 0$ (for $i \ne j$), $k_{ii}=0$
for $i = 1 \cdots N $ and $k_{ii}=1$ for $i = N+1 \cdots 2N $. 
\underline{J}(t) contains the second derivative of the potential as defined
in the earlier paper [8].

We now make two important comments; (i) $\hbar$-scaling leads to a semiclassical
description where the terms of higher powers of $\hbar$ have been left out and 
in the process we identify a stage of early evolution only influenced
by \underline{J} and $\gamma$ (but not by thermal diffusion $D$). (ii) a
key-point in determining the stability of motion rests on the jacobian matrix
(or curvature of the potential) \underline{J} be it regular or chaotic. Since it depends
explicitly on $z_i$ (i. e. , $x_i$ and $p_i$) it is a function of time and not
a constant.

At this point we adapt the theory for the case when the trajectories in 
question are chaotic in nature. Thus, we consider a fully developed strong chaos  
such that the measure of regular region is sufficiently small so that 
${\bf F}(t)$  which is governed by classical 
chaotic  fluctuations in the curvature of the potential can be 
treated as a stochastic process.

Second, we {\it do not make any a priori assumption about the nature of the 
stochastic process ${\bf F}(t)$}. 
The special cases, such as, noise is Gaussian or Markovian or $\delta$-correlated etc.  
have attracted so much attention in the literature that it is necessary to 
emphasize that these approximations have not been made.
Eq.(7) may therefore be regarded as a stochastic 
differential equation with multiplicative noise.

\vspace{0.5cm}

\begin{center}
{\bf B. A Fokker-Planck equation for probability distribution function
of the quantum fluctuations}
\end{center}

Our next task is to find out a differential equation whose average solution 
is given by $\langle \phi \rangle$  
where the stochastic averaging has to be performed over 
the classical noise due to chaos. 
We denote, $\langle \phi \rangle = 
P(\eta,t)$ which defines a probability
distribution of quantum fluctuation variables $\{ \eta \}$ at time t.
To this end we note that 
${\bf F}(t) \cdot {\bf \nabla}$  can be 
partitioned into two parts; a constant part  ${\bf F}_0 \cdot \nabla$      
and a fluctuating part
${\bf F}_1(t) \cdot \nabla$. Thus we write
\begin{equation}
{\bf F} \cdot \nabla = {\bf F}_0 \cdot \nabla + {\bf F}_1 \cdot 
{\bf \nabla} \; \; \; .
\end{equation}

Making use of one of the main results for the theory of linear equation of 
the form (7) with multiplicative noise, we derive an equation for 
$P$ as given by (for details, we refer to [15]);

\begin{eqnarray}
\frac{\partial P}{\partial t} & = &
\left\{ -{\bf F}_0. {\bf \nabla}
+2 N\gamma - \langle {\bf F}_1.{\bf \nabla} \rangle +\int_0^\infty
d\tau \left| \frac{d{\bf \eta}^{-\tau}}{d{\bf \eta}}
\right| \right. \nonumber \\
& & \langle \langle {\bf F}_1({\bf \eta},t) 
.{\bf \nabla}_\tau {\bf F}_1({\bf \eta}^{-\tau}, t-\tau) \rangle 
\rangle \cdot \nabla_{-\tau} \left. \left| \frac{d\eta}{d \eta^{-\tau}}
\right|  \right\} P \; \; \;,
\end{eqnarray}
where $\left| \frac{d \eta}{d \eta^{-\tau}} \right|$ is a jacobian of 
transformation as defined in Refs. [8, 15].

Eq.(9) thus takes into account of two distinct stochastic processes. One is
due to the external reservoir with infinite degrees of freedom which have been 
eliminated and the manybody effect is incorporated through diffusion coefficient
$D$ and dissipative term $\gamma$ in Eq.(4). The other one concerns the 
classical fluctuations of the curvature of the potential \underline{J}
as embodied in ${\bf F}_1$ terms, which is due to classical chaos. While 
the former process is taken into consideration within 
{\it Markovian} description the second process is {\it non-Markovian} 
because of the finite
correlation time $\tau_c$ of classical fluctuations. 
The construction of the associated Fokker-Planck equation is based on 
perturbative cumulant expansion in $\alpha \tau_c$. Following van Kampen [15]
we have assumed that $\tau_c$ is short compared to the average time scale
over which the probability distribution function $P(\eta,t)$ evolves in time.
The convergence of the expansion in $\alpha \tau_c$ thus allows us to retain
upto second order terms and as such one need not go over to higher order
to describe the dynamics. If one takes care of $\tau^2$ terms the theory can
be appropriately extended [15]. We also point out that since we need not 
invoke any a priori approximation on the nature of noise (like Gaussian or
$\delta$ correlated etc.) the values of correlation functions when calculated 
numerically are exact in this sense.

By $\hbar$-scaling one gets
rid of thermal diffusion $D$ in Eq.(9). Thus, the semiclassical description
identifies an early stage of dynamical evolution dominated by dissipation
of the system due to external surroundings and diffusion of fluctuations of the 
curvature of the potential due to classical chaos. The theory developed so far
is valid for N-degree-of-freedom chaotic systems in presence of dissipation.

\vspace{0.5cm}
\begin{center}
{\bf IV. An illustration}
\end{center}
\vspace{0.5cm}
\begin{center}
{\bf A. The Fokker-Planck equation}
\end{center}

We now turn to a simple illustration of the general equation(9) ($N=1$ case)
in terms of 
a low dimensional dissipative chaotic system which allows us to solve the 
equation for probability distribution of quantum fluctuations analytically. 
The model is 
thus  expected to capture some of the essential features of evolution of 
quantum fluctuations in presence of dissipation. We now consider the classical 
motion of a particle of mass m in a potential field $V(x)$ and driven by a
classical field of frequency $\omega_0$. The Hamiltonian is given by
\begin{equation}
H = \frac{p^2}{2 m} + V(x) + gx \cos \omega_0 t  ,
\end{equation}
with $V(x)= a x^4 - b x^2$,
where first and the second terms in Eq.(10) comprise the kinetic and potential
energies of the particle, respectively. 
The third term is the driving term which includes the 
effect of coupling of the system with the field as well as the strength 
of the field.

The classical equation of motion of the particle in presence of damping (at 
a rate $\gamma$) are
\begin{eqnarray}
\dot{x} & = & p \; \; \; , \nonumber \\
\dot{p} & = & - \gamma p -V'(x) - g \cos \omega_0 t \; \; \; .
\end{eqnarray}

Following the method as described in Refs. [8] and [15] the master equation 
(9) in the case of this model chaotic
system can then be written down. This is

\begin{eqnarray}
\frac{\partial P(\eta_1, \eta_2, t)}{\partial t}
& = & \left[ -\frac{\eta_2}{m} \frac{\partial}{\partial \eta_1}
- \left\{ (2b+c+c_2) \eta_1 -2 \eta_2 \gamma \right\} 
\frac{\partial}{\partial \eta_2} \right. \nonumber \\
& + & 2 \gamma + \left. \left\{ c_2 \eta_1^2 
\frac{\partial^2}{\partial \eta_2 \partial \eta_1}+
\eta_1^2 c_1 \frac{\partial^2}{\partial \eta_2^2} -
\frac{\eta_1 \eta_2}{m} c_2 \frac{\partial^2}{\partial \eta_2^2} \right\}
\right] P(\eta_1, \eta_2, t) \; \; ,
\end{eqnarray}

where
\begin{eqnarray}
c & = & \langle \zeta (t) \rangle \; , \nonumber \\
c_1 & = & \int_0^\infty \langle \langle \zeta (t) \zeta (t-\tau) 
\rangle \rangle e^{-2\gamma \tau} d\tau \nonumber \; \; ,\\
c_2 & = & \int_0^\infty \langle \langle \zeta (t) \zeta (t-\tau) 
\rangle \rangle e^{-2\gamma \tau} \tau d\tau \; \; . 
\end{eqnarray}

where $\zeta(t) =12 a x^2 $ represents the fluctuating part of the curvature
of the potential $V(x)$.
The above equation (12) is a Fokker-Planck equation for probability 
distribution  of quantum fluctuations for the model 
chaotic dissipative system.
It is evident that stochastic averaging over classical chaos leads us to the 
above equation and the correlation functions contained in $c, c_1$ and $c_2$. 
The correlation of 
fluctuations of curvature of the classical potential thus determines the 
drift 
and diffusion terms of the Fokker-Planck equation. The appearance of
the variables $\eta_1, \eta_2$
in the diffusion terms precludes the possibility of an exact 
solution of this equation. One thus takes resort to weak noise approximation 
scheme (this is consistent with the assumption that fluctuations are
not too large) under which the diffusion terms are  assumed to be 
constant which are 
given by
\begin{eqnarray}
A' & = & \eta_1^2(0) c_1 - \eta_1(0) \eta_2(0) c_2  \; \; \; , \nonumber \\
B & = & \eta_1^2(0) c_2 \; \; \; .
\end{eqnarray}

Zeroes in  $\eta_1$ and $\eta_2$    
refer to their initial values corresponding to the 
initial preparation of the coherent wave packet. 

We now use the abbreviation
$2b+c+c_2=\omega'^2$ and put $m=1$ for the rest of the treatment. 
The Fokker-Planck 
equation can then be written in a more compact form as follows;
\begin{eqnarray}
\frac{\partial P(\eta_1, \eta_2, t)}{\partial t} & = & \left[-\eta_2
\frac{\partial}{\partial \eta_1}-\omega'^2 \eta_1
\frac{\partial}{\partial \eta_2}  + 2 \gamma +
2 \gamma \frac{\partial}{\partial \eta_2} \eta_2 
\right. \nonumber \\
& + & \left. A' \frac{\partial^2 \langle \phi \rangle}{\partial \eta_2^2}
+ B \frac{\partial^2 \langle \phi \rangle}
{\partial \eta_1 \partial \eta_2} \right] P(\eta_1, \eta_2, t) \; \; \; .
\end{eqnarray}

The above Fokker-Planck equation which governs the evolution of distribution 
of quantum fluctuations $\eta_1$ and $\eta_2$ corresponding to co-ordinate and momentum
variables, respectively, in presence of dissipation has a formal similarity
in structure to Kramers' equation [16] 
which describes the Brownian motion of a
particle in phase space. While the stochasticity in Kramers' equation 
originates from the thermal fluctuations derived from the true statistical 
properties of the reservoir which is a many body system, the stochasticity in the present problem owes 
its origin to the dynamical properties of classical chaos 
in a low dimensional system.

Eq.(15) clearly illustrates three distinct aspects of evolution; (i) in the
absence of $\gamma$ and $A'$ (and $B$) the evolution can be mapped into a 
purely deterministic and reversible Liouville flow under an inverted harmonic
potential $-\frac{{\omega'}^2 {\eta_1}^2}{2}$. Note the $\omega'$ is essentially
$2b$ (which is a parameter in the potential of the Hamiltonian (10)) 
appropriately modified by the average and correlation function of the 
fluctuations of the second derivative of the potential in (10). (ii) the 
reservoir-induced irreversible dissipation ($\gamma$) and (iii) the irreversible 
diffusion ($A'$ and $B$) caused by the fluctuations of the jacobian or 
curvature of the potential (this diffusion is chaotic diffusion and is 
characteristic of the nonlinear system in question and not be confused with 
thermal diffusion $D$ due to external reservoir). The structure of $A'$ and 
$B$ suggests that $B$ in Eqs.(13) and (14) vanishes when $\tau_c \rightarrow 0$, and $B$ can be
identified as a non-Markovian small contribution [15], $A'$ being the 
dominant Markovian part.

We shall now demonstrate in the following two sections (B) and (C) 
the interplay of three above-mentioned aspects of 
evolution in two distinct situations: (a) We first neglect the dissipative term by 
letting $\gamma$ approach zero in Eq.(15) and consider a competitive
effect between the reversible evolution and the chaotic diffusion. To this end we 
closely follow the analysis of Zurek and Paz [17] to establish that there
exists a critical width of Wigner function for quantum noise determined by 
chaotic diffusion $A'$ and $\omega'$. (b) We then take full account of the 
dissipative terms by letting $\gamma$ finite but small to show that how quantum
noise gets amplified by chaotic diffusion in the early stage and then ultimately
settles down to equilibrium under the influence of dissipation. The 
equilibrium is characterized again by a critical width of Wigner function 
determined by chaotic diffusion and dissipation. 
\vspace{0.5cm}
\begin{center}
{\bf B. Reversible evolution of quantum fluctuations and chaotic diffusion}
\end{center}

In order to study the interplay of reversible evolution of quantum fluctuations
(due to Liouville flow expressed through `reversible' operator $L_{\rm rev}$ as shown
below) and chaotic diffusion (expressed in terms of irreversible contribution
$L_{\rm irr}$), we now rewrite the Fokker-Planck Eq.(15) (neglecting the 
$\gamma$-term) in the following form: 
\begin{equation}
\frac{\partial P}{\partial t} = L_k P \; \;,
\end{equation}

where
\begin{equation}
L_k = L_{\rm rev} + L_{\rm irr} \; \; ,
\end{equation}
\begin{equation}
L_{\rm rev}(\eta_1, \eta_2) = - \frac{\partial}{\partial \eta_1} \eta_2 + 
f'(\eta_1) \frac{\partial}{\partial \eta_2}
\end{equation}

and
\begin{equation}
L_{\rm irr}(\eta_1, \eta_2) = A' \frac{\partial^2}{\partial {\eta_2}^2} +
B \frac{\partial^2}{\partial \eta_1 \partial \eta_2}
\end{equation}

Here
\begin{equation}
f(\eta_1) = - \frac{{\omega'}^2 {\eta_1}^2}{2}
\end{equation}

is the inverted parabolic potential, since ${\omega'}^2$ is expressed as
\begin{equation}
{\omega'}^2 = 2b + c + c_2 \; \; ,
\end{equation}

where $2b$ is associated with the saddle point of the quartic potential $V$ in the 
Hamiltonian (10). $\omega'$ is the effective frequency afforded by the unstable
inverted harmonic potential and is dressed by the average (c) and the correlation 
function
($c_2$) of fluctuations of the curvature of the potential $V$. It is 
important
to realize that locally a chaotic flow pattern is similar to that around the 
saddle point from which both the stable and unstable manifolds emanate. The two 
second derivative terms in $L_{\rm irr}$ operator denote the chaotic 
diffusion
of which $A'$ term is the dominant Markovian contribution whereas $B$ is 
a small
non-Markovian addition. In the limit when the fluctuation is fast enough 
such 
that $\tau_c$ approaches zero, $B$ vanishes [15].

To analysis the underlying reversible Liouville flow due to the effective unstable
oscillator, it is now convenient to follow Zurek and Paz [17] and use contracting
and expanding co-ordinates corresponding to stable and unstable directions of
the flow, respectively, as follows:
\begin{eqnarray}
r & = & \eta_2 - \omega' \eta_1 \; \; , \nonumber\\
s & = & \eta_2 + \omega' \eta_1 \; \;.
\end{eqnarray}

The flow generated by Eq.(16) (in absence of diffusion) causes exponential 
contraction in $r$ and expansion in $s$. Expansion in $s$ results in decrease 
in gradient in that direction. We neglect [17] the gradient along that direction 
to obtain the effective evolution of the Wigner function for quantum fluctuations,
after sufficient number of stretching and folding,
\begin{equation}
\frac{\partial P}{\partial t} = L_{\rm eff} (r,s) P
\end{equation}

where
\begin{equation}
L_{\rm eff} (r,s) = \omega'\left(r \frac{\partial}{\partial r}- s \frac{\partial}
{\partial s} + \frac{1}{2} {\sigma_c}^2 \frac{\partial^2}{\partial r^2}\right) \; \;.
\end{equation}

Here ${\sigma_c}^2$, the effective dispersion is given by

\begin{eqnarray}
\sigma_{c}^{2} & = & 2 \left( {\frac{A'}{\omega'}} - B \right)  \nonumber\\
& \simeq & \frac{2A'}{\omega'}  \; \; \; (\rm  for \; B \; small)
\end{eqnarray}

Eq.(23) can be solved by considering the problem as eigenvalue equation [20] for
the operator appearing on the right hand side of (23), i. e. ,
\begin{equation}
L_{\rm eff} P = \lambda P  \; \; \;.
\end{equation}

$L_{\rm eff}$ can be cast into the form of an Hamiltonian for two interacting 
Boson operators [20]. The relevant eigenfunctions are 
$s^n F_m(\frac{r}{\sigma_c})$,
where 
$F_m (x) = \exp (-\frac{x^2}{2}) H_{m-1}$ $(\frac{x}{\sqrt 2})$ and $H_m(x)$ are
Hermite polynomials of order $m$.

The corresponding eigenvalues are $-(n+m+1)\omega'$; $n$ and $m$ being
positive real numbers. (The constant term $\omega'$ in the eigenvalues 
has no dynamical 
significance).

The general solution of Eq.(23) is then given by
\begin{equation}
P(r, s, t)  = \sum_
{\begin{array}{l}
n \geq 0 \\ 
m \geq 1 
\end{array}}
C_{nm} s^n 
F_m(r) \exp- {(n+m)\omega't}
\end{equation}

Taking into consideration [17] that $P$ expands in $s$-direction 
through $s = s_0 \exp{\omega' t}$ and major contribution to the sum (27) can be
attributed to $m = 1$ term we find that $P$ approaches the 
Gaussian with a critical width $\sigma_c$ as follows:
\begin{equation}
P \simeq \frac{1}{\sqrt{2 \pi {\sigma_c}^2}} 
\exp(-\frac{r^2}{2{\sigma_c}^2}) \exp(-{\omega' t}) \nonumber\\
\int_{-\infty}^{+\infty} ds P(r_0, s_0, t=0) 
\end{equation}

It is important to note that the existence of a critical width $\sigma_c$
(expressed as $\frac{2A'}{\omega'}$) reveals an important interplay of 
evolution of quantum fluctuations through reversible dynamics ($\omega'$)
and chaotic diffusion ($A'$) which results due to correlation of fluctuations of
the curvature of the potential. We emphasize once again that the chaotic diffusion coefficient
$A'$ ( and $B$) is not to be confused with the usual thermal diffusion 
coefficient $D$ that appeared in Eq.(4). While the analysis of Zurek and Paz [Z P]
essentially considers a competition arising out of  an interplay of reversible
dynamics of the system and thermal diffusion due to surroundings, the present 
analysis in this section does not take into account of any external influence.
This is an important point of departure from the analysis of Z P.
Both quantum noise around the classical path following the reversible
dynamics and chaotic diffusion due to the correlation of fluctuations 
of the curvature of the
classical potential concern the system itself. It is apparent that the
competition of evolution of quantum fluctuation (in the contracting direction wave
packet gets squeezed) and chaotic diffusion (which causes the spread) as 
inherent in the expression for critical width provides a self-regulatory 
feature in the dynamics. This is characteristic of the chaotic system 
as demonstrated here and is not due to any external influence.

\vspace{0.5cm}

\begin{center}
{\bf C. Evolution of quantum noise and its approach to equilibrium}
\end{center}
\vspace{0.5cm}

We now switch on the dissipative term by letting $\gamma \rightarrow$ finite
but small and show how quantum noise gets amplified by intrinsic classical 
stochasticity
at early stage and then reaches a steady state under the influence of dissipation.
We thus take into consideration of all the three aspects of evolution as 
identified earlier in the full Fokker-Planck equation (15).

We begin by applying the following transformations to fluctuation variables
$\eta_{1}$, $\eta_{2}$ corresponding to position and momentum:
\begin{eqnarray}
\bar{\eta}_{1}& = & \sqrt{\omega} \eta_{1} \; \; \; ,\nonumber \\ 
\bar{\eta}_{2} & = & \frac{1}{\sqrt{\omega}} \eta_{2} \; \; \; .
\end{eqnarray}

We then let
\begin{eqnarray}
\beta & = & \bar{\eta}_{1}+i\bar{\eta}_{2}   \; \; \; , \nonumber \\
\beta^{*} & = & \bar{\eta}_{1}-i\bar{\eta}_{2} \; \; \; .
\end{eqnarray}

Then the appropriate transformation of derivatives in Eq.(15) yields the following 
Fokker - Planck equation corresponding to Eq. (15).
\begin{eqnarray}
\frac{\partial p}{\partial t} & = &
\left(-\frac{\beta \omega}{2i}+\frac{\beta^*\omega}{2i}-
\frac{i \omega'^2 \beta}{2\omega}-\frac{i\omega'^2 \beta^*}{2\omega}
+\gamma \beta - \gamma \beta^* \right) 
\frac{\partial p}{\partial \beta} \nonumber \\
& + & \left(-\frac{\beta \omega}{2i}+\frac{\beta^*\omega}{2i}+
\frac{i \omega'^2 \beta}{2\omega}+\frac{i\omega'^2 \beta^*}{2\omega}
-\gamma \beta + \gamma \beta^* \right) 
\frac{\partial p}{\partial \beta^*} \nonumber \\
& + & 2A \frac{\partial^2 p}{\partial \beta^* \partial \beta}  + 2\gamma
+(-A+iB)\frac{\partial^2 p}{\partial \beta^2}+
(-A-iB) \frac{\partial^2 p}{\partial {\beta^*}^2} \; \; ,
\end{eqnarray}

where for notational convenience we have used 
\begin{eqnarray}
A & = & \frac{A'}{\omega} \; \; \; ,\nonumber \\
P(\bar{\eta_{1}},\bar{\eta_{2}},t)
& = & p(\beta,\beta^{*},t) \; \; \; .                                                                   
\end{eqnarray}

We now search for the Green's function or conditional probability solution 
for the system at $\beta$ and $\beta^{*}$ at time $t$ given that it had the 
values $\beta'$ and $\beta'^{*}$ at $t=0$. The initial condition which is 
required to bring forth quantum-classical correspondence is represented by
\begin{equation}
p(\beta, \beta^*, t=0) = 
\frac{\epsilon}{\pi} e^
{-\epsilon (\beta-\beta^*) (\beta'-\beta^{*'})}
\end{equation}

which corresponds to a coherent state.
We then look for a solution of the equation (31) of the form
\begin{equation}
p(\beta, \beta^*, t) | \beta', \beta^{*'},0) = e^{G(t)},
\end{equation}

where
\begin{equation}
G(t)=-\frac{1}{\Gamma(t)}[\beta-\Omega(t)][\beta^*-\Omega(t)]
+{\rm ln}\nu(t) \; \; .
\end{equation}

$G(t)$ is determined in terms of the time-varying parameters $\Gamma(t)$,
$\Omega$(t) and $\nu$(t) which follow a set of ordinary differential equations
given in Appendix A. The important relevant quantity required for the present
analysis is $\Gamma$(t) which is given by,

\begin{equation}
\Gamma(t) = \Gamma(0)e^{-2\gamma t}+\frac{A}{\gamma}(1-e^{-2\gamma t}),
\end{equation}

The Green's function or the conditional probability solution (34) can then be
employed to calculate the various theoretical quantities of which, the 
uncertainty in cordinate $\Delta \eta_1$ and that in momentum are obtained
as follows,

\begin{eqnarray}
\Delta \eta_1^2 & = & \langle \eta_1^2 \rangle - \langle \eta_1 \rangle^2
=\frac{1}{\omega} \left[ \frac{\Gamma(t)}{2} \right]  \; \; \; ,\\
\Delta \eta_2^2 & = & \langle \eta_2^2 \rangle - \langle \eta_2 \rangle^2
= \left[ \frac{\Gamma(t)}{2} \right] \omega \; \; \; .
\end{eqnarray}

We are led to two important results of this section.

(i) The uncertainty product
$\Delta \eta_1 \Delta \eta_2$ at any time is given by
\begin{equation}
\Delta \eta_1 \Delta \eta_2 = \frac{1}{2} \Gamma(t)
\end{equation}

where $\Gamma(t)$ is determined by Eq.(36) 
subject to  initial conditions (A2). This implies that we choose 
$\epsilon=1$ to satisfy the minimum uncertainty 
product condition for t=0, for the wave packet.
Eq.(39) relates the evolution of 
quantum noise as a function of time in terms of 
$\Gamma(t)$ which by the 
virtue of Eq.(36) is determined by the initial condition $\Gamma(0)$ 
[Eq.(A2)]
and the other two parameters A and $\gamma$. Note that A is the chaotic diffusion 
coefficient defined by Eqs.(14) and (32) [this is not to be confused with the thermal
diffusion coefficient $D$ in Eq. (3) which arises due to the
interaction with the
surroundings] and $\gamma$ refers to the dissipation rate of the system   
in contact with the surroundings. Since A ( and $A'$) is related to the fluctuations of 
the curvature of the classical potential $\zeta(t)$ through $c_{2}$ and
$c_{1}$ in Eq.(14), the origin of diffusion coefficient A is essentially the
classical chaos. Eq.(39) thus illustrates how the initial quantum noise gets 
amplified by classical stochastisity.

(ii) The equilibrium condition is governed by the long time limit of the 
conditional probability function
\begin{equation}
P_{ss} = \lim_{t \rightarrow \infty}
P(\bar{\eta}_1, \bar{\eta}_2, t |
\bar{\eta}_1', \bar{\eta}_2',0) \; \; \; .
\end{equation}

This is given by
\begin{equation}
P_{ss}
= \bar{\nu} \exp \left[ - \frac{(\bar{\eta}_1^2+\bar{\eta}_2^2)}{A/\gamma}
\right] \; \; \; ,
\end{equation}

where
\begin{eqnarray*}
\bar{\nu} = \frac{\Gamma(0) \nu(0) \gamma}{A} \; \; \; .
\end{eqnarray*}

Furthermore, the equilibrium condition implies that there exist a critical limit
to the expansion of phase space. This is apparent from the uncertainty product
relation (39) as follows;
\begin{equation}
\Delta \eta_1 \Delta \eta_2 |_{t \rightarrow \infty} = \frac{A}{\gamma} \; \; .
\end{equation}

The existence of critical limit which also appears as a width of equilibrium
distribution (41) is a consequence of competition between chaotic diffusion
which attempts to expand the wave packet and dissipation $\gamma$ which has the
opposite tendency and ultimately leads to a compromise steady state.

The relation (39) thus illustrates the effect of classical chaos on quantum 
fluctuations at the semiclassical level. In order to examine the initial 
divergence of quantum variances it is thus necessary to calculate the 
correlation functions in $c_{1}$ and $c_{2}$ numerically by solving the 
classical equations of motion (11) with specific initial conditions which 
admit chaos. In order to allow ourselves a fair comparison with fully 
quantum calculation and verify the theoretical propositions, we shall return
to this issue in the next section.

\vspace{0.5cm}
\begin{center}
{\bf V. \hspace{0.2cm} Numerical Calculations; Classical and Quantum}
\end{center}

\vspace{0.5cm}
To analysis the growth of quantum fluctuations quantitatively 
[Eq. 39] we now 
consider the dissipative classical chaotic motion governed by Eq.(11). We
choose the parameter values m=1, a=0.5, b=10. $\omega_{0}$=6.07. The coupling 
constant cum field strength, g and the damping rate $\gamma$ 
are the parameters 
which have been varied from set to set. 
These two quantities essentially determine the two competeting processes in the dynamics,
e. g. , the fundamental strong coherent interaction ($g$) between the double-well
oscillator and the external field, (strong coupling being responsible for 
classical chaos) and the irreversible weak decay of the oscillator ($\gamma$).
Keeping in mind the approximations involved we thus consider a situation where
$g >> \gamma$ (while scaled g varies over a range 12-18, the variation of scaled $\gamma$
is of the order $10^{-4}$-$10^{-1}$). In quantum mechanical terms the 
situation is somewhat reminiscent of a coherent regime in typical cavity
quantum optical problem where the Rabi frequency far exceeds the damping rate
(a typical Rabi frequency $\sim 10^{12}$/sec compared to damping rate $\sim 10^9$/sec).
The parameter space in the dissipation-free version of the model is chosen 
from Lin and Ballentine [18].
We choose the initial condition 
$x_{0}$=-3.5 and $p_{0}$=0.0, which ensures strong global chaos [18]. 
Note that $c_1$ and $c_2$ as expressed in Eq.(13) are the integrals over the
correlation function of $\zeta(t)$ [ $\zeta(t)$ is the fluctuating part of the
second derivative of the potential $V(x)$ and is given by $\zeta(t)=
-12 a x^2$]. To calculate the correlation function 
$\langle \langle \zeta(t) \zeta(t-\tau) \rangle \rangle$ and the average $\langle \zeta(t) \rangle$ it is
necessary to determine long time series in $\zeta(t)$ by numerically solving 
the classical equations of motion (11) for $x$ and $p$. The next step is to 
carry out averaging over the time series. Since $\langle \zeta(t) \rangle$ and
$\langle \langle \zeta(t) \zeta(t-\tau) \rangle \rangle$ are classical 
quantities to be calculated exactly and the underlying model is nonintegrable,
one must have to take resort to numerical integration. Any approximation in this
regard amounts to making specific assumption about the nature of the classical
noise. We have already pointed out in Sec. III that a main virtue of the 
treatment of stochastic differential equations with multiplicative noise by 
van Kampen [15] is that one does not have to make any a priori assumption
about the nature of the noise $\zeta(t)$. 
However it is essential that the time of decay of the correlation of 
classical quantities like $\langle \langle \zeta(t)\zeta(t-\tau) \rangle \rangle$
must not be too long (evidently the treatment of weak chaos which involves
long correlation time is out of space of the present theory). The theoretical footing of our 
analysis thus remains intact in numerical calculation.

Once the average quantities like $\langle \zeta(t) \rangle$ and
$\langle \langle \zeta(t) \zeta(t-\tau) \rangle \rangle$ are known, our next task
is to calculate the integrals over time $\tau$ [in Eq.(13)] to determine 
$c_1$ and $c_2$.
One of the primary considerations of the theory described in the earlier 
section is that one takes care of classical fluctuation of $\zeta(t)$ upto 
second order such that the correlation time     
$\tau_c$ is short but finite. On a coarse-grained 
time scale over which the distribution function P($\eta$, t) proceeds, the 
process is approximately Markovian. Numerical implementation of
this near-Markovian character in our analysis rests on the fact that we 
consider the first fall of the correlation function to determine the 
cut off time for time integration  in $c_1$ and $c_2$ . With these numerical values of $c_{1}$ and
$c_{2}$ one obtains A and $\Gamma(t)$ of Eq.(39). In Figs 1 and 2 we plot 
(dotted curve) $ln \Delta \eta_1 \Delta \eta_2$ 
as a function of time $t$ for different values of $g$ and 
$\gamma$. 
 
For a full quantum-mechanical calculation to verify the basic 
theoretical propositions of semiclassical dynamics, we now return to Eq. (3).
The eigenvectors     
$\{| n \rangle \}$ of a harmonic oscillator which satisfies 
$(\hat{p}^2/2m + (1/2) m \omega^2 \hat{x}^2) | n \rangle =
[(n+1/2)\hbar \omega] | n \rangle$  
are chosen as basis vectors to solve Eq. (3). The 
frequency $\omega$ is arbitrarily 
adjusted to economize the size of the basis 
set. For the present purpose we choose  
$\omega=6.25$, $\hbar=1$, and 120
basis vectors. In this representation 
the equation of motion for the reduced 
density [Eq. (3)] matrix elements is given by
\newpage
\begin{eqnarray}
\frac{d\rho_{nm}}{dt} & = & -i/\hbar
\left[ \sum_k H_{nk} \rho_{km} - \sum_l \rho_{nl} H_{lm} \right] 
\nonumber \\
& + &
\frac{\gamma}{2} \left\{ \sqrt{(n+1) (m+1)} \rho_{n+1, m+1} -
(n+m) \rho_{nm} \right\} 
\nonumber \\
& + & D \left\{ \sqrt{(n+1) (m+1)} \rho_{n+1, m+1} - (n+m+1)
\rho_{nm} \right\} \; \; \; .
\end{eqnarray}

Here $H_{mn}$ is as given in Ref[18]. $H$ is given by Eq. 10.

The Eq.(43) describes the evolution of both population 
(diagonal elements) and 
coherence (off-diagonal elements). 
Without making any approximation regarding the 
separation of time scales for evolution of them, 
we have carried out numerical
solution of 120x120 equations 
for density matrix elements of the reduced system 
as a typical initial value problem. We follow  time scale of 
the period of driving force                        
$T=2\pi/\omega_0$, so that $t= \tau T$ , where $\tau$ is a dimensionless
quantity.

Quantum-classical correspondence 
is maintained through construction of minimum
uncertainty wave packets         
$| \alpha_{x, p} \rangle$ of Gaussian form in position and momentum 
representations having position    
$\langle x \rangle$ and average momentum     
$\langle p \rangle$ such that
\begin{equation}
\langle \alpha_{x, p} | n \rangle = \left[ \exp(-0.5 | \alpha |^2 )
\right] \frac{\alpha^n}{\sqrt{n!}} \; \; \; , 
\end{equation}
where,
\begin{eqnarray*}
\alpha = \sqrt{m \omega/2} [ \langle x \rangle + (i/m\omega) \langle p 
\rangle ]  \; \; \; .
\end{eqnarray*}

The quantum evolution is followed by locating the average position and average
momentum of the initial wave packet corresponding to the initial position and 
momentum of a classically chaotic trajectory. As a numerical check 
we have compared our results 
with those of Lin and Ballentine [18] in classical and 
quantum cases for D=0, $\gamma$=0. Another important check for the 
numerical 
calculation is to keep $Tr \rho = 1$ for the entire evolution.

Following Eq.(39) we plot the variation of 
$ln [\Delta \eta_1 \Delta \eta_2]$ ($\Delta \eta_1$ and $\Delta \eta_2$              
are the quantum variances 
corresponding to position and momentum, respectively) as a 
function of time for several values of $g$ but for a fixed $\gamma$ (0.001) 
in Figs 1(a-c). It has already been pointed out 
that the major input for the theoretical 
quantity is the chaotic diffusion coefficient A which is further related to 
$c_{1}$ and $c_{2}$ , i.e., to classical correlation function of the 
curvature of the potential. The theoretical curves are 
denoted in 
Figs 1(a-c) by the dotted lines. It is evident that after a sharp initial
growth the uncertainty product tends to settle down to some final value. The
initial rate becomes large with the increase of $g$. The variation of rate 
constants with $g$ is shown in table I. The theoretical analysis has been 
supplemented by the fully quantum mechanical calculations based on numerical 
integration of Eq.(43) by launching Gaussian wave packets centered around
the classical position ($x_{0}$) and momentum ($p_{0}$) corresponding to the
chaotic trajectory as described in the earlier part of this section. These
numerical curves have been superimposed in Figs 1(a-c) for the corresponding
values of $g$ and $\gamma$
(An initial flat plateau region has been cut off to make the
rise more prominent [8]).
It may be observed that the agreement between
the theoretical and numerical curves is quite satisfactory so far as the
initial growth part is concerned. The agreement is  better for larger 
g-values.
Table I also gives a relative comparison between the theoretical and 
numerical estimate of the rate constants.

The effect of relaxation rate $\gamma$ is analyzed 
in Figs 2(a-c) where we have 
plotted $ln(\Delta \eta_1 \Delta \eta_2)$          
as a function time for a fixed $g$ but for several values
of $\gamma$. It is easy to see that 
the attainment of equilibrium significantly 
depends on $\gamma$. For low values of $\gamma$ 
it takes larger time to attain 
the equilibrium state. To have a rough estimate of the time required to 
achieve equilibrium  we have determined the time in which the
$d ln(\Delta \eta_1 \Delta \eta_2)/dt$      
reduces to $\approx 1\%$  of its initial value. 
These times have been tabulated 
in the table II  for the theoretical and numerical curves in 
Figs 2(a-c). 
It is also apparent from the figures that with the increase of 
$\gamma$ (within weak damping limit) the time for attainment of equilibrium gets shorter and one also 
finds a better agreement between the theoretical and numerical curves. 

We thus observe an interesting interplay of chaotic diffusion and dissipation 
through $A$ and $\gamma$. 
These are the two major factors that determine the initial 
rate of divergence (Eq.39) and approach to equilibrium (Eq.41) of quantum 
variances in a system which is classically chaotic. Since at very low values 
of $\gamma$, the approach to equilibrium gets 
slowed down significantly, one has 
to wait for a very long time for the steady state. Now if the time becomes 
too long then dispersion of the wave packet becomes too much 
so that the validity of a
semiclassical analysis becomes questionable, where one takes 
into consideration of terms of lowest 
order in $\hbar$. At this point higher order derivative terms containing
the potential in Eq.(4) 
become significant. Thus with the higher values of $\gamma$ 
(within weak damping limit) and 
$g$ one finds very good agreement between semiclassical theory and numerical 
analysis(e.g., Fig 2c).

Another pertinent point need be mentioned here. Apparently from the methods 
used one should expect that the agreement between quantum and semiclassical
results improves as the damping is decreased. However in Fig.2 the opposite
seems to be the case. To be more precise we first note that for smaller values of 
damping and short time both the results agree well. The discrepency arises, 
at smaller values damping and in the long time regime. 
It must be understood that the full quantum simulation works in the entire
time range of dynamics and is valid for weak dissipation. On the other
hand the semiclassical analysis, although can be extended to large damping
limit is not valid for a very long time regime since quantum effects begin to 
dominate. However, when damping is increased the time required to attain the  
equilibrium gets
shorter and one finds a better agreement between semiclassical and quantum 
analysis (since one need not  carry out the calculation of dynamics over
a larger length of time when higher order quantum effects take over).

\vspace{0.5cm}
\begin{center}
{\bf VI. \hspace{0.2cm}Summary of the main results and conclusions}
\end{center}

In this paper we have considered the quantum evolution of  
dissipative, chaotic systems whose classical limit is chaotic. 
Making use of appropriate $\hbar$-scaling 
(analogous to van Kampen's  $\Omega$-expansion) of 
the equation for Wigner's 
quasiprobability distribution 
function which takes into account of dissipation and 
thermal diffusion terms on the basis of a system - 
reservoir theory, we derive 
a semiclassical dynamical equation for probability distribution function for   
quantum fluctuations. The equation incorporates dissipation due to 
surroundings and fluctuations of the curvature of the classical potential 
(arising out of classical chaos) in addition to standard Liouville flow terms 
as the essential features of the semiclassical dynamics.
Appropriate treatment of fluctuations of the curvature of the potential
which is amenable to a theoretical analysis of multiplicative noise for short 
but finite correlation time leads us to a Fokker-Planck equation where the 
drift and diffusion terms have their origin in the intrinsic dynamical 
properties of the classical chaotic system as well as the dissipation of the 
system  due to its contact with the surroundings. Formally, the equation is 
identical in structure with Kramers' equation [16] which describes a Brownian
dynamics in phase space demonstrates the interplay of three distinct aspects 
of evolution (a) purely deterministic reversible Liouville flow (b) irreversible
chaotic diffusion which is intrinsic to the system itself and (c) irreversible
dissipation due to coupling of the system to external surroundings. We have
corroborated our semiclassical analysis by numerical simulation of the full
quantum operator master equation.

We now summarize the main conclusions of this study.

(i) $\hbar$-expansion and a stochastic treatment of the curvature of the 
classical potential in terms of a $\tau_c$-expansion identify a specific stage of quantum evolution with three
distinct aspects in an open system described by a Fokker-Planck equation. 
These are (a) reversible deterministic Liouville flow (b) irreversible chaotic
diffusion which is intrinsic characteristic of the nonlinear system itself 
(c) irreversible dissipation of the system induced by external reservoir. This 
stage of evolution is less likely to be affected by thermal diffusion which
is otherwise primarily responsible for decoherence processes.

(ii) Chaotic diffusion although an intrinsic property of the system imparts 
{\it a kind of irreversibility} in the dissipation-free evolution which has a truly deterministic
origin.

(iii) In the dissipation-free case, an interplay of approximately reversible
Liouville flow and chaotic diffusion sets a {\it critical limit} on the width
of Wigner function for quantum noise undergoing time evolution.

(iv) Our results show how the initial quantum noise gets amplified by chaotic
diffusion and then ultimately equilibrated with the passage of time under the
influence of dissipation.

(v) We establish that there exists a {\it critical limit to the expansion 
of phase space}. This is determined by an interplay of chaotic diffusion and dissipation
and has an important bearing on the evolution of entropy of an open system.

The present analysis is based on $\hbar$ -expansion and cumulant expansion
in $\tau_c$. Both being convergent perturbative schemes promise suitable 
extension to higher orders to reveal more details of subtlities of quantum 
evolution and long time memory effects due to weak classical chaos. Appropriate
extension of the theory in this direction is worth-pursuing in future.

\noindent
{\bf Acknowledgments:} 
B. C. Bag is indebted to the Council of Scientific and
Industrial Research for partial financial support. D. S. Ray is thankful
to the Department of Science and Technology for partial financial support. 

\newpage

\begin{center}
{\bf Appendix A \\
The solution for conditional probability(47)}
\end{center}

We are to see that, by suitable choice of
$\Omega(t)$, $\Omega^*(t)$, $\Gamma(t)$ and $\nu(t)$ 
Eq.(31) can be solved subject to the initial condition
\begin{eqnarray*}
p(\beta, \beta^*, 0) | \beta', \beta^{*'},0) = 
\frac{\epsilon}{\pi} e^
{-\epsilon (\beta-\beta') (\beta^*-\beta^{*'})} \;.
\hspace{6.5cm}{\rm A1} 
\end{eqnarray*}

Comparison of this with (34) with G(0) shows that
\begin{eqnarray*}
\Gamma(0) = \frac{1}{\epsilon} \; \; , \; \; 
\Omega(0) = \beta' \; \; , \; \;
\Omega^*(0) = \beta^{*'} \; \; , \; \; \nu(0)=\frac{\epsilon}{\pi} \; \; .
\hspace{4.5cm}{\rm A2} 
\end{eqnarray*}

If we put (34) in Eq.(31) and equate the coefficients of equal powers of $\beta$
and $\beta^{*}$ we obtain after some algebra the  following 
set of equations
\begin{eqnarray*}
\frac{1}{\Gamma^2} \frac{d \Gamma}{dt} & = & -\frac{2\gamma}{\Gamma} +
\frac{2A}{\Gamma^2} \; \; , \hspace{8.8cm}{\rm A3} \\
\frac{d\Omega}{dt} & = & \left[ \frac{\omega}{2i}-\frac{i\omega'^2}{2\omega}
-\gamma + \frac{2(A-i B)}{\Gamma}\right] \Omega^*  \nonumber \\
& + & \left[ \frac{\omega}{2i}+\frac{i\omega'^2}{2\omega}-\gamma 
\right] \Omega \; \; , \hspace{7.2cm}{\rm A4} \\
\frac{1}{\nu} \frac{d\nu}{dt} & = & 2 \gamma -\frac{2A}{\Gamma}
= -\frac{1}{\Gamma} \frac{d\Gamma}{dt} \; \;, \hspace{7.5cm}{\rm A5} 
\end{eqnarray*}

together with the conjugate of $d\Omega/dt$       
equation. The relevant solutions of these which
satisfy the initial conditions above and are necessary for the present
analysis are seen to be
\begin{eqnarray*}
\Gamma(t) = \Gamma(0)e^{-2\gamma t}+\frac{A}{\gamma}(1-e^{-2 \gamma t}),
\hspace{7.9cm}{\rm A6} 
\end{eqnarray*}

and
\begin{eqnarray*}
\nu(t) = \frac{\Gamma(0) \nu(0)}
{\Gamma(0)e^{-2 \gamma t} + \frac{A}{\gamma}(1-e^{-2\gamma t})}.
\hspace{7.9cm}{\rm A7} 
\end{eqnarray*}

The solution of $\Omega$(t) can be similarly obtained after appropriate 
algebraic manipulations. Having known $\Gamma(t)$, 
$\nu(t)$, $\Omega(t)$ and $\Omega^{*}(t)$ one calculates the conditional
probability function for quantum fluctuations given by Eq.(34).

We now calculate the quantum fluctuations of position and momentum 
variables. Since the conditional probability $P$ 
is given 
by Eq.(34), this
together with (35) and (32) may be employed to calculate first and second
moments. Thus we express
\vspace{0.5cm}
\begin{eqnarray*}
\langle \bar{\eta}_1 \rangle = \frac{
{\int \int}_{-\infty}^\infty
P(\bar{\eta}_1, \bar{\eta}_2, t | \bar{\eta}'_1, \bar{\eta}'_2,0)
\bar{\eta}_1 d\bar{\eta}_1 d\bar{\eta}_2}
{{\int \int}_{-\infty}^\infty
P(\bar{\eta}_1, \bar{\eta}_2, t | \bar{\eta}'_1, \bar{\eta}'_2,0)
d\bar{\eta}_1 d\bar{\eta}_2}
\hspace{6cm}{\rm A8} 
\end{eqnarray*}

in terms of conditional probability $P$ using (30) and (34). Explicit 
calculation yields
\begin{eqnarray*}
\langle \bar{\eta}_1 \rangle = Re [\Omega(t)] \; \; \; ,
\hspace{9.9cm}{\rm A9} 
\end{eqnarray*}

where $\Omega(t)$ is a solution of (A4). Similarly we obtain
\begin{eqnarray*}
\langle \bar{\eta}_1^2 \rangle = \frac{1}{2} \Gamma(t) + 
{[Re \{\Omega(t)\}]}^2 \; \; \; .
\hspace{8cm}{\rm A10} 
\end{eqnarray*}

The conjugate variable to $\bar{\eta}_{1}$ is $\bar{\eta}_{2}$ whose 
average is given by
\begin{eqnarray*}
\langle \bar{\eta}_2 \rangle = Im[\Omega(t)] \; \; \; .
\hspace{9.9cm}{\rm A11} 
\end{eqnarray*}

Similarly
\begin{eqnarray*}
\langle \bar{\eta}_2^2 \rangle = \frac{1}{2} \Gamma(t)+
{\{Im[\Omega(t)]\}}^2 \; \; \; .
\hspace{8cm}{\rm A12} 
\end{eqnarray*}
The above expressions (A9-A12) for the averages can then be utilised to calculate
the uncertainties in coordinate and momentum variables as given in Eq. (39).

\newpage
\begin{center}
{\bf REFERENCES}
\end{center}

\vspace{0.5cm}

\begin{enumerate}
\item See, for example, 
W. H. Louisell, Quantum Statistical Properties of Radiation
(Wiley, New York, 1973).
\item A. O. Caldeira and A. J. Leggett, Physica \underline{121A}, 587 (1983); 
A. J. Leggett and A. O. Caldeira, 
Phys. Rev. Letts. \underline{46}, 211 (1981). 
\item G. Gangopadhyay and D. S. Ray, J. Chem. Phys. \underline{46}, 4693 
(1992) ;
F. Haake, H. Risken, C. Savage and D. F. Walls, Phys. Rev. \underline{A34}, 
3969 (1986).
\item A. Tameshtit and J. E. Sipe, Phys. Rev. \underline{A47}, 1697 (1993).
\item T. Dittrich and R. Graham, Ann. Phys.  \underline{200}, 363 (1990); Z. Phys. B
 \underline{62}, 515 (1986); R. Graham, Phys. Rev. Letts. \underline{53}, 
2020 (1984); D. Cohen, Phys. Rev. \underline{A44}, 2292 (1991).
\item R. F. Fox and T. C. Elston, Phys. Rev. \underline{E49}, 3683 (1994).
\item L. Bonci, R. Roncaglia, B. J. West and P. Grigolini, Phys. Rev. Letts.
\underline{67}, 2593 (1991).
\item S. Chaudhuri, G. Gangopadhyay and D. S. Ray, Phys. Rev. \underline{E54}, 
2359 (1996).
\item S. Chaudhuri, G. Gangopadhyay and D. S. Ray, Phys. Rev. \underline{E47}, 
311 (1993).
\item S. Chaudhuri, G. Gangopadhyay and D. S. Ray, Phys. Rev. \underline{E52}, 
2262 (1995).
\item A.K. Pattanayak and P. Brumer, Phys. Rev. Letts, \underline{79}, 4131 (1997)
\item T. Dittrich, B. Oelschlaegel and P. H\"anggi, Euro. Phys. Letts. 
\underline{22}, 5 (1993).
\item B. Sundaram and P. W. Milonni, Phys. Rev. \underline{E51}, 1971 (1995).

\item R. F. O'Connell and E. P. Wigner, Phys. Lett. \underline{85A}, 121 (1981).
\item N. G. van Kampen, Phys. Rep. \underline{24}, 171 (1976).
\item P. H\"anggi, P. Talkner and M. Borokovec, Rev. Mod. Phys. \underline{ 62}, 251 (1990).
\item W. H. Zurek and J. P. Paz, Phys. Rev. Letts. \underline{72}, 2508 (1994).
\item W. A. Lin and L. E. Ballentine, Phys. Rev. Letts. \underline{65},  2927 (1990).
\item S. Chaudhuri, D. Majumdar and D. S. Ray, Phys. Rev.\underline{E56}, 
5816 (1996).
\item H. Risken, The Fokker-Planck Equation (Springer- Verlag, Berlin, 1989).
\end{enumerate} 

\newpage
{\bf Table I}. Comparison of the rate constant 
of initial divergence of uncertainty
product calculated numerically (from fully quantum considerations,
Eq. 43),
$k_{{\rm numerical}}$, with that calculated theoretically,
$k_{{\rm theoretical}}$, [from Eq. (39)].

\vspace{0.5cm}
\begin{center}
\begin{tabular}{|c|c|c|}
\hline
\hline
$g$ & $k_{{\rm numerical}}$ & $k_{{\rm theoretical}}$\\
\hline
12 & 1.94 & 1.93 \\
14 & 2.44 & 2.48 \\
18 & 3.02 & 3.08 \\
\hline
\end{tabular}
\end{center}

\vspace{1cm}
{\bf Table II.} Comparison of the time required to reach equilibrium
calculated numerically (from fully
quantum considerations, Eq. 43), $t_{{\rm numerical}}$, with that
calculated theoretically, $t_{{\rm theoretical}}$.

\vspace{0.5cm}
\begin{center}
\begin{tabular}{|c|c|c|}
\hline
\hline
$\gamma$ & $t_{{\rm numerical}}$ & $t_{{\rm theoretical}}$ \\
\hline
0.001 & 10.5 & 12.5 \\
0.01 & 8 & 10 \\
0.09 & 5.3 & 6.3 \\
\hline
\end{tabular}
\end{center}

\newpage

\begin{center}
{\bf Figure Captions}
\end{center}

\begin{enumerate}
\item Fig.1 Plot of log of uncertainty  product with time 
(time scale corresponds to the time period of the driving 
force $\frac{2\pi}{\omega_0}$)
for different values of $g$. The continuous line represents
the numerical calculation (fully quantum). The dotted line refers to
semiclassical calculation ( Eq.(39)). (a) $g$ = 12.0, 
(b) $g$ = 14.0 and (c) $g$ = 18.0.
( Both units are arbitrary).
\item Fig.2 Plot of log of uncertainty  product with time (time scale 
corresponds to the time period of the driving force $\frac{2\pi}{\omega_0}$)
for different values of $\gamma$. The continuous line represents
the numerical calculation (fully quantum). The dotted line refers to
semiclassical calculation ( Eq.(39)). (a) $\gamma$ = 0.0005, 
(b) $\gamma$ = 0.005 and (c) $\gamma$ = 0.045.
( Both units are arbitrary).
\end{enumerate}

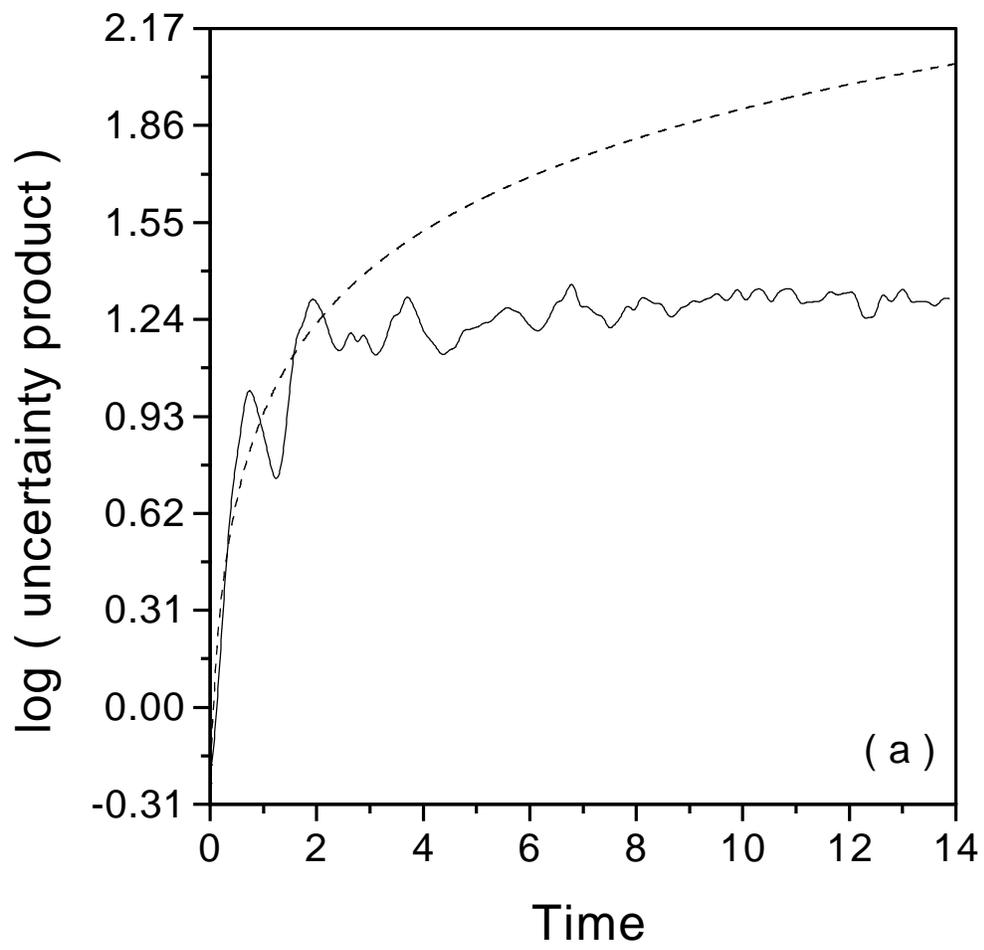
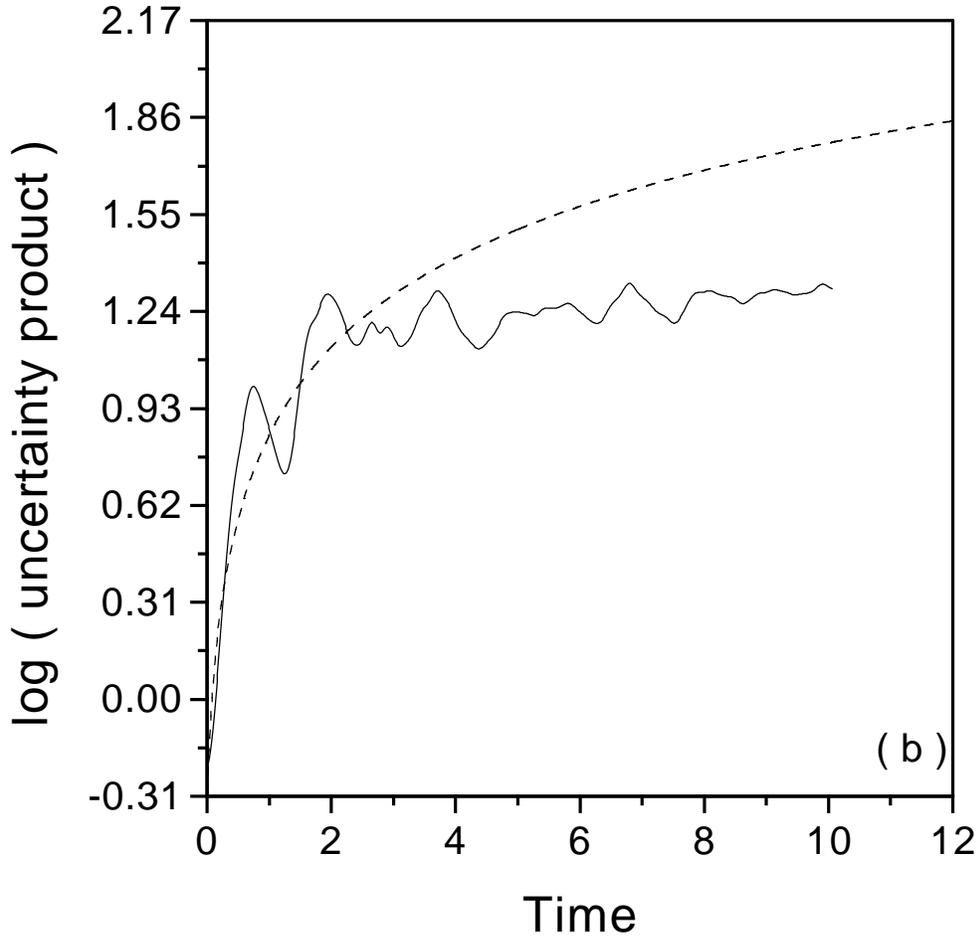
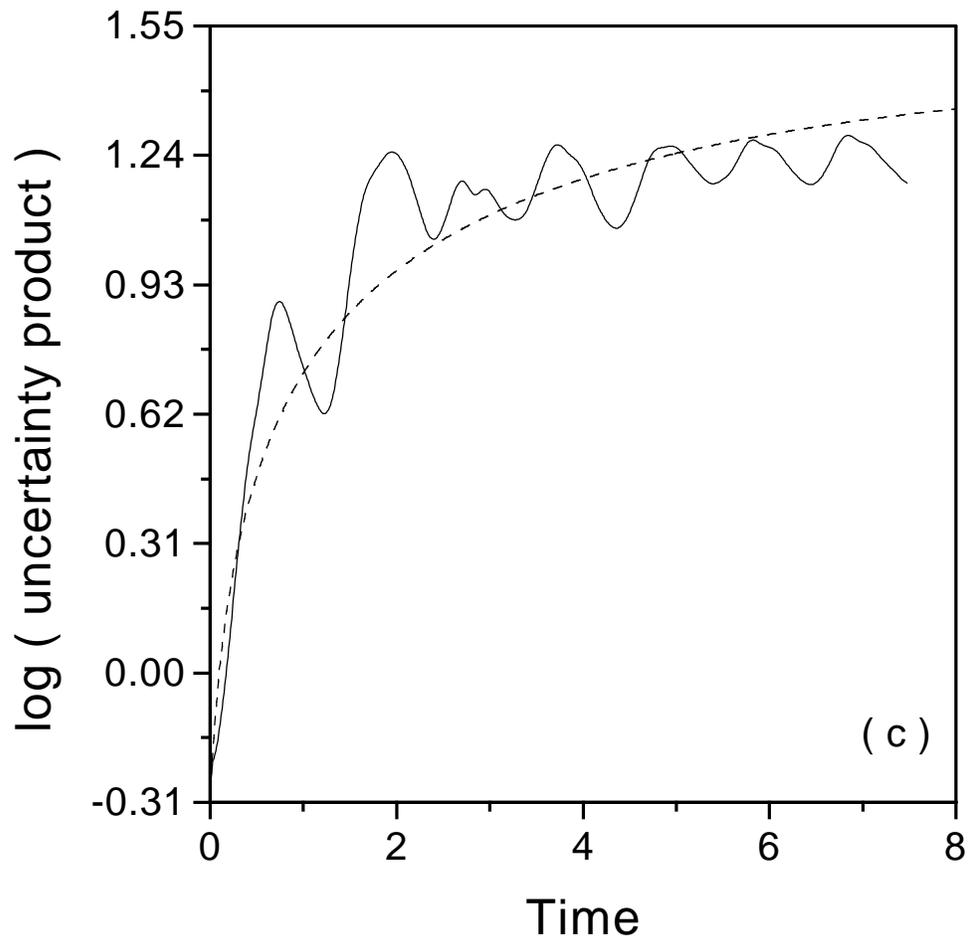
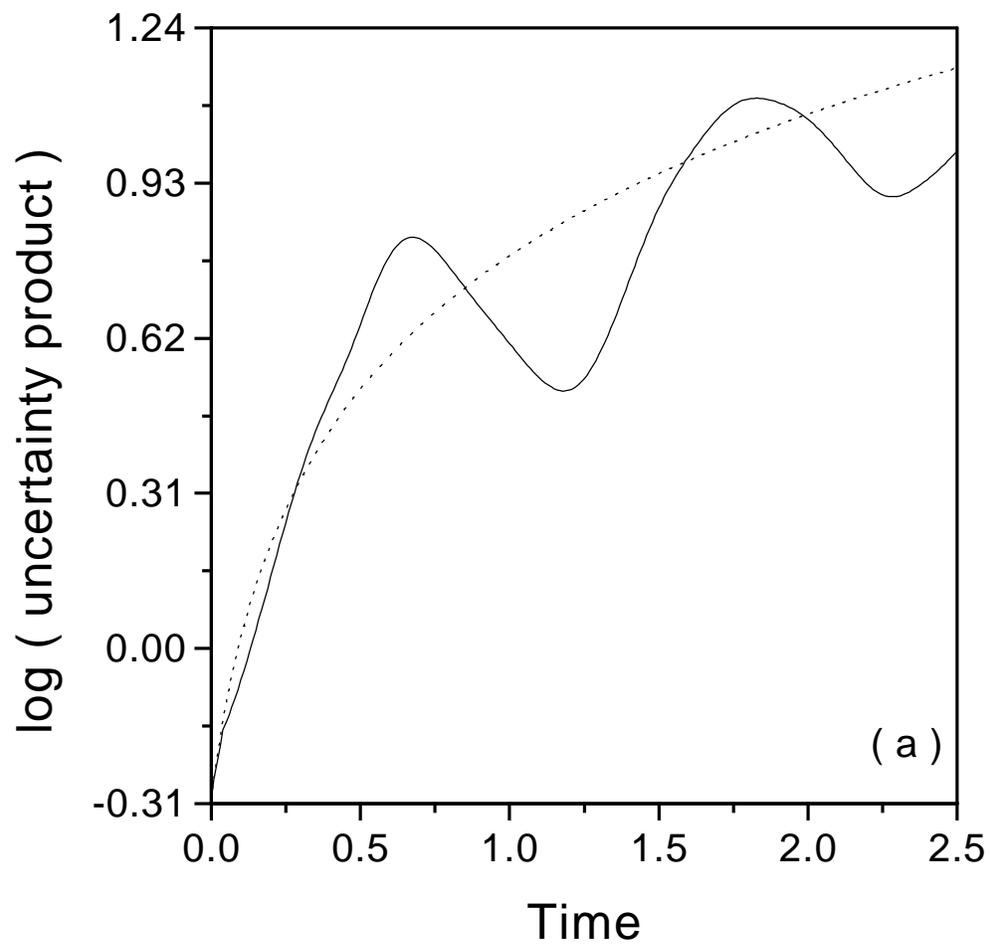
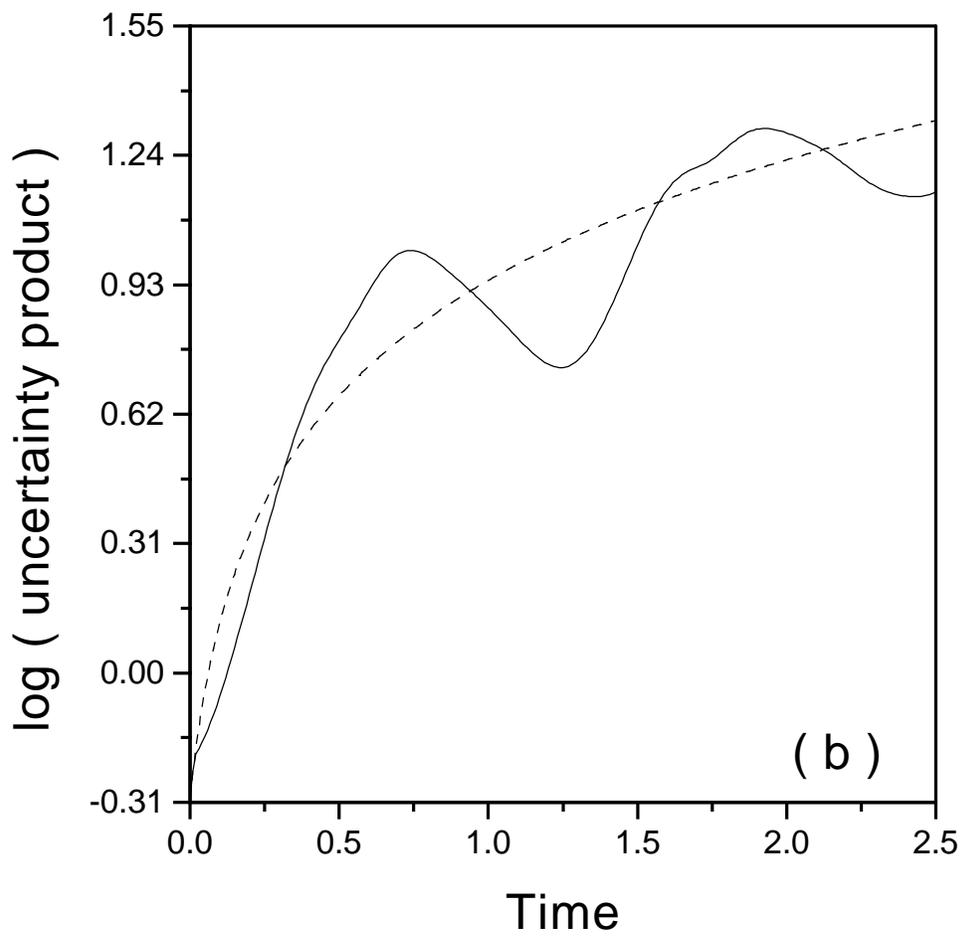
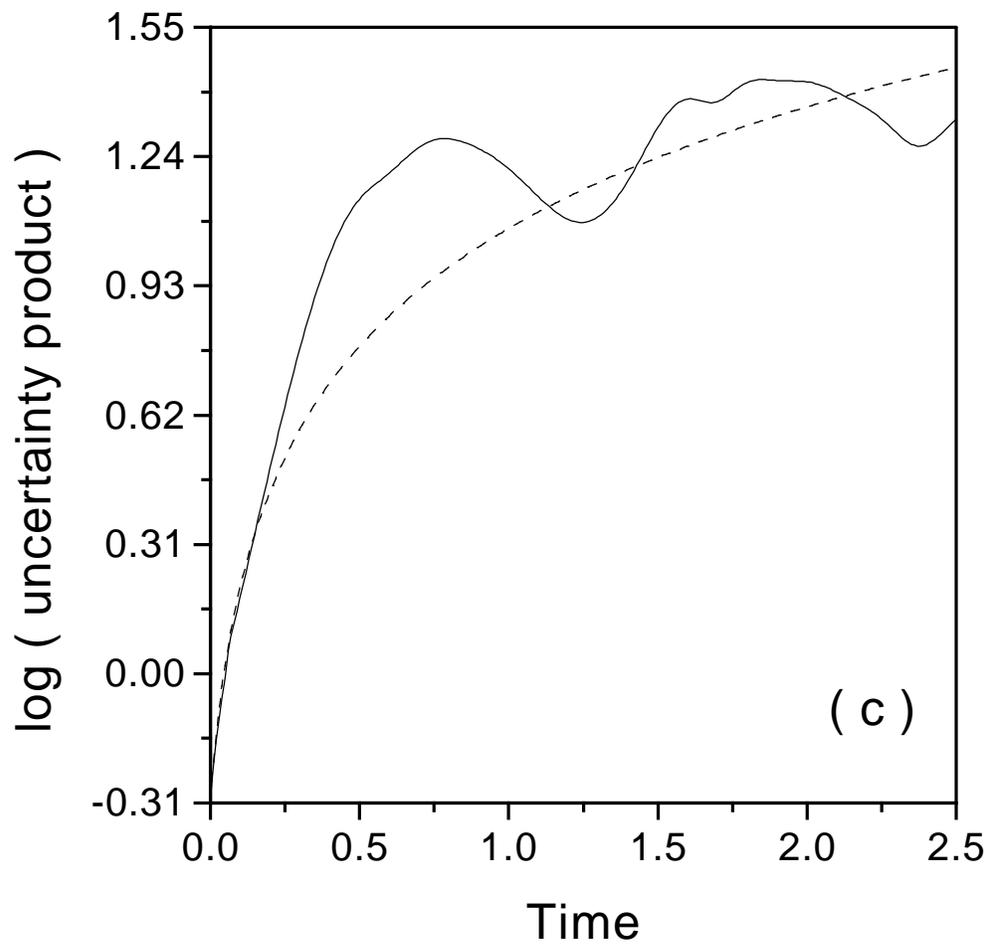
\end{document}